\documentclass{aa}
\usepackage{upgreek}
\usepackage{graphicx}
\usepackage{subcaption}
\usepackage{threeparttable}
\usepackage[varg]{txfonts}
\usepackage{natbib}
\usepackage{float}
\usepackage{soul}
\newlength{\linwx}
\usepackage{color}

\begin{document}

\title{How drifting and evaporating pebbles shape giant planets III: The formation of WASP-77A b and $\tau$ Bo\"otis b}

\author{
Bertram Bitsch \inst{1}, Aaron David Schneider\inst{2,3}, Laura Kreidberg \inst{1}
}
\offprints{B. Bitsch,\\ \email{bitsch@mpia.de}}
\institute{Max-Planck-Institut f\"ur Astronomie, K\"onigstuhl 17, 69117 Heidelberg, Germany
  \and
  Niels Bohr Institutet, Københavns Universitet, Blegdamsvej 17, 2100 København, Denmark
  \and
  Institute of Astronomy, KU Leuven, Celestijnenlaan 200D, 3001, Leuven, Belgium
}
\abstract{Atmospheric abundances of exoplanets are thought to constrain the planet formation pathway, because different species evaporate at different temperatures and thus radii in the protoplanetary disc, leaving distinct signatures inside the accreted planetary atmosphere. In particular the planetary C/O ratio is thought to constrain the planet formation pathway, because of the condensation sequence of H$_2$O, CO$_2$, CH$_4$, and CO, resulting in an increase of the gas phase C/O ratio with increasing distance to the host star. Here we use a disc evolution model including pebble growth, drift and evaporation coupled with a planet formation model that includes pebble and gas accretion as well as planet migration to compute the atmospheric compositions of giant planets. We compare our results to the recent observational constraints of the hot Jupiters WASP-77A b and $\tau$ Bo\"otis b. WASP-77A b's atmosphere features sub-solar C/H, O/H, H$_2$O/H with slightly super-solar C/O, while $\tau$ Bo\"otis b's atmosphere features super-solar C/H, O/H and C/O with sub-solar H$_2$O/H. Our simulations qualitatively reproduce these measurements and show that giants like WASP-77A b should start to form beyond the CO$_2$ evaporation front, while giants like $\tau$ Bo\"otis b should originate from beyond the water ice line. Our model allows the formation of sub- and super-solar atmospheric compositions within the same framework. On the other hand, simulations without pebble evaporation, as used in classical models, can not reproduce the super-solar C/H and O/H ratios of $\tau$ Bo\"otis b's atmosphere without the additional accretion of solids. Furthermore, we identify the $\alpha$ viscosity parameter of the disc as a key ingredient regarding planetary composition, because the viscosity drives the inward motion of volatile enriched vapor, responsible for the accretion of gaseous carbon and oxygen. Depending on the planet's migration history through the disc across different evaporation fronts, order-of-magnitude differences in atmospheric carbon and oxygen abundance should be expected. Our simulations additionally predict super-solar N/H for $\tau$ Bo\"otis b and solar N/H for WASP-77A b. We conclude thus that pebble evaporation is a key ingredient to explain the variety of exoplanet atmospheres, because it can explain both, sub- and super-solar atmospheric abundances.
}

\authorrunning{Bitsch et al.}\titlerunning{The formation of WASP-77A b and $\tau$ Bo\"otis b}\maketitle

\section{Introduction}
\label{sec:Introduction}

Planet formation models are mostly constrained by the observed mass, radius and orbital distance distributions of exoplanets and their corresponding occurrence rates (e.g. \citealt{2008ApJ...673..487I, 2009A&A...501.1139M, 2014A&A...565A..96G, 2017ASSL..445..339B, 2018MNRAS.474..886N, 2018AJ....156...24M, 2020MNRAS.493.1013A}). However, current and future observations will expand on the constraints on planet formation by adding atmospheric abundances to the data. The link of atmospheric abundances to planet formation models are mostly discussed via the planetary C/O ratio (e.g. \citealt{2011ApJ...743L..16O, 2017MNRAS.469.4102M, 2017MNRAS.469.3994B, 2020A&A...642A.229C, 2021A&A...654A..71S, 2022arXiv220413714M}), which changes with distance to the host star due to condensation of different carbon and oxygen bearing species at different disc temperatures (e.g. H$_2$O, CO$_2$, CH$_4$, CO). Additionally, nitrogen has been discussed as a potential tracer for the formation location of exoplanets \citep{2019AJ....157..114B, 2021ApJ...909...40T}, but also for Jupiter \citep{2019A&A...632L..11B, 2019AJ....158..194O, 2021A&A...654A..72S}.

Detailed observation of atmospheric abundances of exoplanets are still quite rare compared to the bulk of observed exoplanets, but with increasing measurement precision we are beginning to see a diversity in atmospheric properties. Some planets appear to have sub-solar abundances of water (e.g. \citealt{2017MNRAS.469.1979M, 2020AJ....160..280C}), whereas others are metal-rich (e.g. \citealt{2018AJ....155...29W}). Even though there is substantial scatter in the mass-atmospheric metallicity relation, there may be an overall tendency for hot Jupiters to be water-poor \citep{2019ApJ...887L..20W}. On the other hand, it was suggested that the super-stellar alkali metal abundance of some of these hot Jupiters \citep{2019ApJ...887L..20W} might be consistent with inward migration and accretion of planetesimals rich in refractories but poor in water ice \citep{2022MNRAS.509..894H}. However, this process ignores that large amounts of volatiles and even evaporated refractories could be accreted via the gas phase, depending on the migration history of the giant planet \citep{2017MNRAS.469.3994B, 2021A&A...654A..71S, 2021A&A...654A..72S}. Additionally \citet{2006MNRAS.367L..47G} suggested that a large fraction of the heavy element content of giant planets could be accreted at the late stages of the disc evolution, where photoevaporation mainly removes hydrogen and helium from the disc, leading to a natural enrichment of heavy elements. However, a general application of this theory would have difficulties explaining sub-solar compositions. Furthermore, outer giant planets could block inward flowing pebbles,  depleting the inner discs of volatiles and pebbles \citep{2016Icar..267..368M, 2021A&A...649L...5B, 2021A&A...654A..71S} and thus altering the composition of growing planets (e.g. \citealt{2021A&A...649L...5B}).

Recent observations of exoplanet atmospheres were able to not only constrain single molecules precisely, but also derive C/H, O/H, C/O and H$_2$O/H within the planetary atmospheres with great precision. \citet{2021Natur.598..580L} observed sub-solar C/H, O/H and H$_2$O/H with a slightly super-solar C/O in the atmosphere of the 1.8 $M_{\rm Jup}$ inflated ($1.2R_{\rm Jup}$) hot Jupiter WASP-77A b, orbiting its host star in 1.36 days. In contrast, \citet{2021AJ....162...73P} using SPIRou/CFHT reported super-solar C/H and O/H with slightly super-solar C/O in combination with a sub-solar water abundance in the atmosphere of the 6 Jupiter mass hot Jupiter $\tau$ Bo\"otis b, which orbits its host star in 3.3 days. Observing the same planet, \citet{2022arXiv220514975W} reported a near solar water abundances of the same planet using observations via NIR CARMENES, clearly indicating that the water abundance of $\tau$ Bo\"otis b is still debated. We note that these observations pertain to the day side of these planets, implying that the abundances may be affected by disequilibrium processes linked to zonal flows and strong day/night side temperature variations (e.g. \citealt{2002A&A...385..166S}). Furthermore, interior processes and chemical reactions in the planetary atmosphere might influence the water abundance in the upper atmosphere (e.g. \citealt{2015ApJ...813...47M, 2021MNRAS.505.5603B}). As a consequence, it is difficult to draw definite conclusions on planet formation just from the water abundance alone, which is why we mostly rely on C/H and O/H to derive conclusions for our planet formation model.

Even though evidence seems to indicate that the atmospheric abundance is not a tracer of the bulk abundances (e.g \citealt{2021ExA...tmp...56H, 2022arXiv220504100G}), also considering recent constraints from Jupiter \citep{2017GeoRL..44.4649W, 2018A&A...610L..14V, 2019ApJ...872..100D, 2022arXiv220301866M} and Saturn \citep{2021NatAs...5.1103M}, we nevertheless adopt, for simplicity, the assumption that the atmospheric abundance is a tracer of the bulk abundances. In this work we focus on planet formation simulations in discs governed by pebble growth, drift and evaporation to study the atmospheric abundances of growing and migrating planets. In addition to the orbital parameters and planetary masses, we focus specifically to match the atmospheric constraints of WASP-77A b \citep{2021Natur.598..580L} and $\tau$ Bo\"otis b \citep{2021AJ....162...73P}, because these planets represent the two extreme ends of the exoplanet population: sub- and super-solar abundances, where both extremes have to be matched within the same planet formation scenario. Furthermore their formation is not only constrained through the C/O ratio, but also through C/H, and O/H, not available for most other observed exoplanets, giving the highest level of constraints to planet formation models. In the following we present a planet formation model that can explain sub- and super-solar atmospheric abundances without invoking the additional accretion of solids into planetary atmospheres, as required by classical models where pebble evaporation is not taken into account \citep{2011ApJ...743L..16O}.

\section{Planetary growth model}
\label{sec:growth}

The planet formation model we use is described in detail in \citet{2021A&A...654A..71S}. In particular the model includes pebble growth and drift \citep{2012A&A...539A.148B}, pebble evaporation and condensation at ice lines \citep{2021A&A...654A..71S}, planet growth via pebble \citep{Johansen2017} and gas accretion \citep{2021MNRAS.501.2017N} as well as planet migration \citep{2011MNRAS.410..293P, 2015Natur.520...63B}. The initial planetary mass is set by the pebble transition mass, at which the planet starts to accrete efficiently from the Hill regime \citep{2012A&A...544A..32L}. During the build up of the planetary core we attribute 10\% of the accreted pebbles to a primordial heavy element atmosphere. The planet switches to gas accretion once it has reached its pebble isolation mass \citep{2014A&A...572A..35L, 2018arXiv180102341B}, at which the planet opens a small gap in the protoplanetary disc, preventing further pebble drift interior to the planet. At this stage the planet can only accrete a gaseous component (incl. H, He and volatiles), but is unable to accrete any solids. This accretion picture is fundamentally different to an accretion scenario including planetesimals, which could still be accreted into planetary atmospheres once the planet starts to accrete an envelope (e.g. \citealt{2018NatAs...2..873A}), giving rise to abundance differences in refractories and volatiles in planetary atmospheres \citep{2021A&A...654A..72S}.

In contrast to the classical step-function picture of the gas phase C/O \citep{2011ApJ...743L..16O}, the evaporation of inward drifting pebbles allows more extreme C/O ratios in the gas phase of the disc (e.g. \citealt{2021A&A...654A..71S}), where the C/O ratio over the whole disc radius can vary over a few orders of magnitude depending which carbon or oxygen rich materials evaporate (e.g. the C/O ratio is strongly sub-solar just interior of the water ice line due to the evaporation of water ice) and how fast the vapor moves through the disc. Furthermore, this effect allows super-solar abundances of oxygen and carbon in the gas phase, unachievable in a model without pebble evaporation.

\begin{figure}
 \centering
 \includegraphics[scale=1.0]{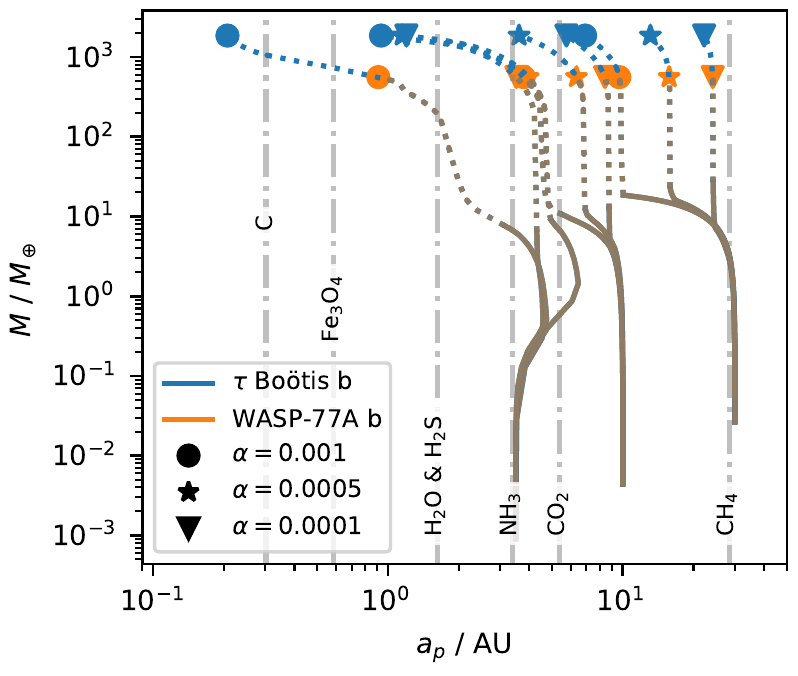}
 \caption{Growth tracks of planets starting at different orbital positions in discs with different viscosities (marked by different symbols). The solid lines mark solid accretion, while the dashed lines mark gas accretion. The dots mark the final masses of WASP-77A b (orange) and $\tau$ Bo\"otis b (blue). The evaporation lines for different species are shown for $\alpha=5 \times 10^{-4}$ and their distance to the host star increases with increasing $\alpha$ parameter, but do not evolve in time in our model for simplicity. We note that the growth of the $\tau$ Bo\"otis b analogues are a continuation of the growths of the planets resembling WASP-77A b, indicated by a change in color for the corresponding growth tracks. The planetary growth is then stopped once the mass of WASP-77A b or $\tau$ Bo\"otis b is reached.
   \label{fig:growth}
   }
\end{figure}

We investigate here the growth and migration of planetary embryos starting at three different locations (3.5, 10 and 30 au) in discs with three different $\alpha$ viscosity parameters ($\alpha=10^{-4}, 5\times10^{-4}, 10^{-3}$). We change the initial planetary positions and $\alpha$ viscosity parameters in section~\ref{sec:alpha}. We further assume a solar composition for the different chemical elements \citep{2009ARA&A..47..481A} with a solar dust-to-gas ratio ($\epsilon_0=0.0124$), motivated by the measured near solar abundances of WASP-77A, \citep{2021arXiv211202031K}.

Here we follow a model where 60\% of all carbon is locked in refractories \citep{2021A&A...654A..72S}, motivated by ISM carbon abundances \citep{2015PNAS..112.8965B}. As a consequence, 20\% of all carbon grains contribute to CO, 10\% to CH$_4$ and 10\% to CO$_2$. We do not include a chemical evolution model because the drift time-scales are shorter than the chemical reaction time-scales \citep{2019MNRAS.487.3998B}. We note that the chemical composition of the material that the planet accretes depends strongly on the composition of the material in the disc, which is normally assumed to be linked to the stellar abundances. We will investigate the effects of varying stellar abundances on planetary compositions in a future work. We use the standard disc parameters from our previous simulations \citep{2021A&A...654A..71S, 2021A&A...654A..72S}, shown in Table~\ref{tab:parameters} and discussed in appendix~\ref{ap:parameters}.

In Fig.~\ref{fig:growth} we show the growth tracks of nine planets in their corresponding discs with varying $\alpha$ and initial orbital position. The planetary growth and migration is strongly influenced by the disc's viscosity, because the viscosity sets the migration rate and when the planets are able to open gaps\footnote{Lower viscosities allow early gap opening, resulting in an earlier transition to the viscously driven type-II migration regime.} \citep{2007MNRAS.377.1324C}. Additionally the viscosity sets the gas accretion rates once the gap is opened and the planet can only accrete what the disc can provide \citep{2020A&A...643A.133B, 2021MNRAS.501.2017N}. As a result planets in discs with higher viscosities grow larger and migrate more compared to their counterparts in discs with lower viscosities. The initial outward migration of the planets starting at 3.5 au is driven by the heating torque, which acts efficiently due to the fast accretion of pebbles \citep{2015Natur.520...63B, 2020arXiv200400874B}. During their migration, the planets cross different evaporation fronts and can then start to accrete the corresponding evaporated material with the gas. 

Starting from a given orbtial position we integrate until the final planetary masses of WASP-77A b and $\tau$ Bo\"otis b have been reached. We make the assumption that the protoplanetary disc dissipates once the final planetary mass has been reached. This implies that the formation of $\tau$ Bo\"otis b takes longer than the formation of WASP-77A b, because $\tau$ Bo\"otis b needs to accrete more material. This results in typical disc lifetimes between 1.5 to 4 Myr, depending on the planet and on the disc's viscosity (see above).

\section{Atmospheric abundances of giant planets}
\label{sec:atmosphere}

\begin{figure}
 \centering
 \includegraphics[scale=1.0]{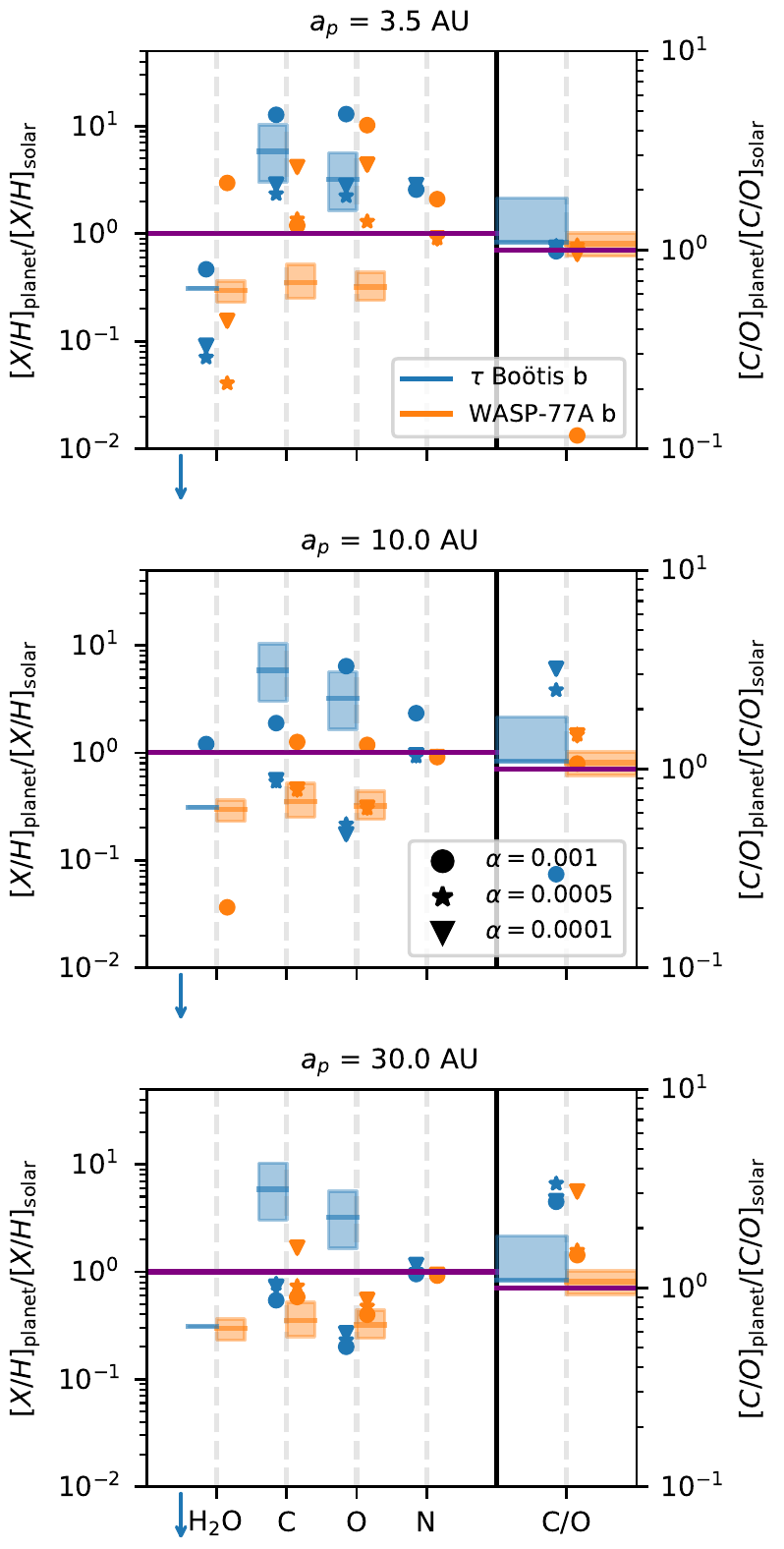}
 \caption{Atmospheric abundances of volatiles normed to solar for the different planets shown in Fig.~\ref{fig:growth} at the end of their evolution. The initial planetary position increases from top to bottom. The colors mark the different planets, while the different symbols mark the different levels of viscosity. The orange bands mark the measured atmospheric abundances of WASP-77A b \citep{2021Natur.598..580L}, while the blue band marks the constraints for $\tau$ Bo\"otis b, where H$_2$O/H is less than $10^{-2}$ \citep{2021AJ....162...73P}, as indicated by the blue arrow. The slightly sub-solar water measurements of \citet{2022arXiv220514975W} are marked with the horizontal blue bar. Some of our simulations feature water abundances below $10^{-2}$ and are thus not shown in the figure. Please note the different scale for C/O.
   \label{fig:owen}
   }
\end{figure}

In Fig.~\ref{fig:owen} we show the atmospheric abundances of the planets shown in Fig.~\ref{fig:growth}. In particular we focus on C, O, N as well as on the C/O and the water abundance in the planetary atmospheres, even though our simulations also track other elements \citep{2021A&A...654A..71S, 2021A&A...654A..72S}. We particularly include the water abundances, because observations seems to indicate that hot Jupiters harbor tendentially a sub-solar water abundance \citep{2019ApJ...887L..20W}. The water abundance in Fig.~\ref{fig:owen} has been calculated using the chemical equilibrium interpolator of petitRADTRANS \citep{2020A&A...640A.131M}, which determines the chemical abundances assuming that the atmosphere is in chemical equilibrium. The water abundance in such a model is then only dependent on the temperature, the pressure and the elemental composition of the planet.
We calculate the water abundances of $\tau$ Bo\"otis b and WASP-77A b by using an average over a pressure range from $10^{-4}$ to 10 bar. For the temperature, we assume that both planets are on tidally locked orbits with $a_p$ = 0.024 au and $a_p$ = 0.046 au for WASP-77A b and $\tau$ Bo\"otis b respectively, leading to equilibrium temperatures of 2366K and 2314K respectively. Using the double gray analytical pressure-temperature profile of \citet{2010A&A...520A..27G} with an interior temperature of $T_{\rm int}$ = 200K, an infrared opacity of $\kappa_{\rm IR} = 0.01$ cm$^2$/g and an optical opacity of $\kappa_{\rm vis} = 0.004$ cm$^2$/g, we obtain (for reference) a temperature at $10^{-3}$ bar of 2145K and 2097K for WASP-77b and $\tau$ Bo\"otis b respectively. The elemental composition which is used to calculate the water abundance with the chemical equilibrium code is consistently inferred from the C/O ratio and the heavy element content of the atmospheres in our planet formation models. We discuss how the lack of pebble evaporation would influence the planetary compositions in Appendix~\ref{ap:noevap}.

Within our simulations, some very clear trends emerge. Planets accreting most of their material in the inner disc region, have super-solar C/H and O/H values, mostly due to the accretion of water and carbon grain vapor, which enriches the gas to super-solar values (in contrast to simulations without pebble evaporation, see Appendix~\ref{ap:noevap}). Planets migrating interior to the water ice line feature super-solar water abundances, explaining the large water abundances of some observed giant exoplanets without any problems \citep{2018AJ....155...29W}. However further inward migration across the carbon grain evaporation line increases the planetary C/O, preventing efficient water formation in the atmosphere. The increase of the planetary C/H in the inner disc, strongly depends on the amount of carbon grains in the disc.

Planets staying mostly in the outer disc harbor lower C/H and O/H values, which can even be sub-solar, especially if the planets do not migrate across the CO$_2$ or H$_2$O ice lines\footnote{Migration across the water evaporation front does not immediately imply an efficient accretion of water vapor, because the pebbles are trapped in the pressure bump exterior to the planet and can only evaporate once the planet migrated far enough that the pebbles in the pressure bump can evaporate.}. As a consequence, these planets accrete mostly CO and CH$_4$ gas (besides N$_2$), resulting in super-solar C/O ratios. Only once the planets migrate across the CO$_2$ ice line is a sub-solar C/O ratio possible. Additionally, these outer planets all feature a sub-solar water abundance in their atmosphere, in line with observed exoplanets \citep{2019ApJ...887L..20W}. These exoplanets could form in the outer disc and then be scattered inwards, where they are observed now.

Our simulations predict solar to super-solar nitrogen contents in the giant planets, where the nitrogen content is larger if planets migrate across the NH$_3$ evaporation front. The simulation that matches the C/H and O/H for $\tau$ Bo\"otis b best ($a_{\rm P}=3.5$ au, $\alpha=10^{-4}$), features an N/H content with a similar level of enrichment. In contrast, the simulations that matches C/H and O/H of WASP-77A b best ($a_{\rm P}=10$ au, $\alpha=10^{-4}$ or $\alpha=5 \times 10^{-4}$), feature solar N/H. The results for the simulations with different viscosities are similar because both WASP-77A b analogues do not migrate across the CO$_2$ evaporation front.

\section{Influence of the disc's viscosity}
\label{sec:alpha}

\begin{figure*}
 \centering
 \includegraphics[scale=1.0]{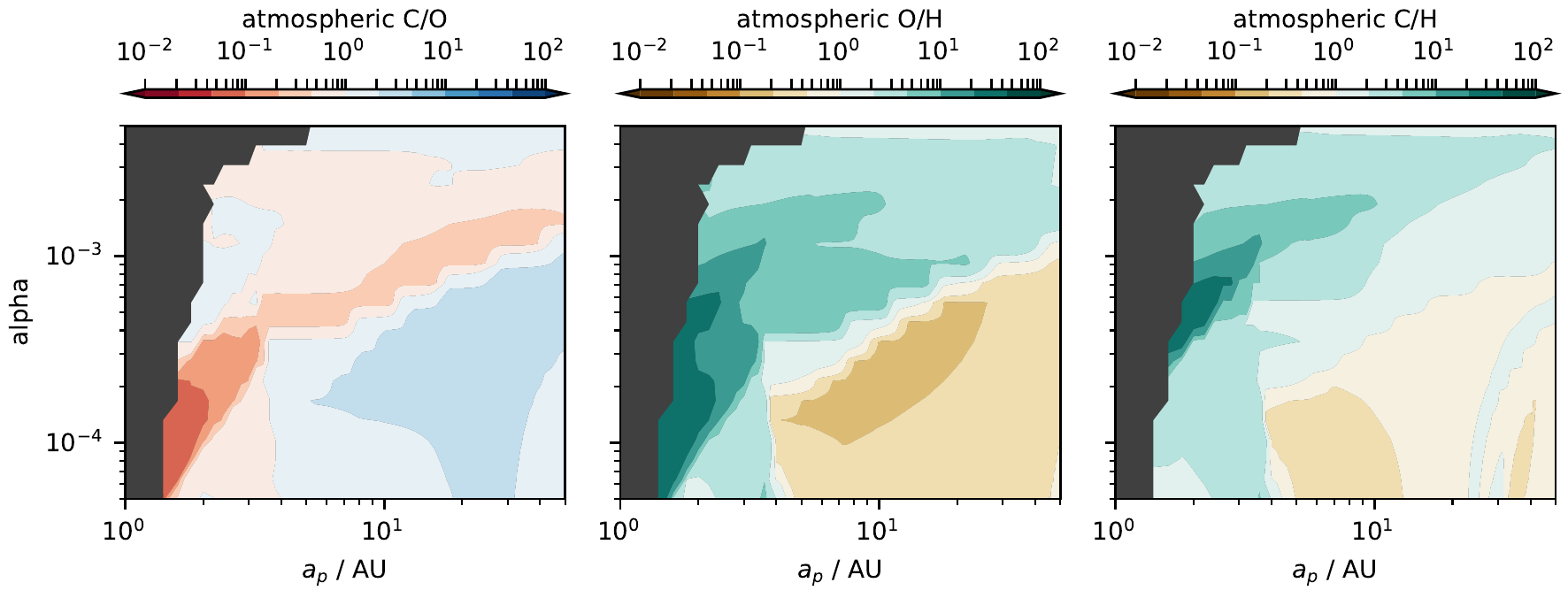}
 \caption{Atmospheric C/O, O/H and C/H of planets forming in discs with different $\alpha$-viscosity at different initial positions after 3 Myr of evolution. The dark regions correspond to planets that have not reached masses above 100 Earth masses. We note that planets growing in discs with higher viscosities migrate farther and are generally more massive than planets growing in discs with lower viscosities \citep{2021A&A...654A..71S}.
   \label{fig:alpha}
   }
\end{figure*}

The viscosity in the protoplanetary disc determines the disc's evolution \citep{1974MNRAS.168..603L}, but it is also important for the composition of planetary atmospheres, because the viscosity sets how fast the vapor, originating from evaporating pebbles, can move inwards. For example, at low viscosity, the CO and CH$_4$ vapor released at their specific evaporation front will only reach the inner edge of the disc at very late times, while this vapor will reach the inner disc regions within a Myr at high viscosity \citep{2021A&A...654A..71S}.

To stress the importance of the viscosity on the atmospheric abundances, we show in Fig.~\ref{fig:alpha} the atmospheric abundances of planets growing in our model at different initial positions in disc's with different $\alpha$-viscosity parameter, ranging from $5 \times 10^{-5}$ to $5 \times 10^{-3}$. We show the atmospheric C/O, C/H and O/H of these planets after 3 Myr of disc evolution. As the final planetary masses of the planets in these simulations do not match those of WASP-77A b and $\tau$ Bo\"otis b, we do not mark the atmospheric constraints for these planets in Fig.~\ref{fig:alpha}.

Our simulations clearly show a dependency of the atmospheric composition on the disc's viscosity, in contrast to earlier models without pebble evaporation \citep{2011ApJ...743L..16O}). In particular sub-solar O/H and C/H atmospheric values are only possible in the outer disc regions at low viscosities ($\alpha \leq 10^{-3}$). On the other hand, super solar C/H and O/H ratios are possible in the outer disc at higher viscosities due to the more efficient inward diffusion of CO and CH$_4$ gas. In the inner disc regions, super-solar values of C/H and O/H are easily possible, especially interior to the CO$_2$ and H$_2$O evaporation fronts. Here the exact value of the C/H and O/H ratio also depends on the disc's viscosity, where larger O/H values are possible at lower viscosity, because the smaller viscosity prevents efficient removal of the water vapor (see \citealt{2021A&A...654A..71S} for a discussion on the disc's water content).

The atmospheric C/O ratio reveals the already established trends from simpler models without evaporation \citep{2011ApJ...743L..16O, 2017MNRAS.469.4102M} that the atmospheric C/O increases with increasing distance to the central star at low viscosity. However, the evaporation of inward drifting pebbles allows more extreme C/O ratios compared to the simple model (see also Appendix~\ref{ap:noevap}), where C/O ratios below 0.1 and above 4-5 are possible if evaporation is included. However the parameter space for these extreme values is limited to low viscosities, where the initial pile-up of vapor is larger \citep{2021A&A...654A..71S}.

\section{Model limitations}
\label{sec:discussion}

Recent advances in high resolution observations have enabled precise constraints on the abundances of both carbon- and oxygen-bearing molecules (e.g. \citealt{2017AJ....153...83B, 2019ApJ...887L..20W, 2021Natur.592..205G, 2021AJ....162...73P, 2021Natur.598..580L}). Here we explore the success of our model in explaining the measured abundances for two case studies: WASP-77A b \citep{2021Natur.598..580L} and $\tau$ Bo\"otis b \citep{2021AJ....162...73P}. In Fig.~\ref{fig:owen} we show the abundances constraints from WASP-77A b and $\tau$ Bo\"otis b, where our model is able to qualitatively reproduce these very different abundance measurements, especially in low viscosity environments.

Our model is able to produce sub-solar C/H and O/H ratios with a nearly solar C/O ratio for planets forming completely exterior to the CO$_2$ evaporation front in line with the observed abundances of WASP-77A b \citep{2021Natur.598..580L}. The inferred super-solar abundances of $\tau$ Bo\"otis b \citep{2021AJ....162...73P} are also reproduced in our model, but they require a formation/migration interior to the CO$_2$ evaporation line, indicating a formation closer to the host star. We stress that our formation scenario is only based on the constraints of C/H, O/H and C/O and not on the water abundances itself, because it is not very well constraint at this point \citep{2021AJ....162...73P, 2022arXiv220514975W} and can also be influenced by interior processes.

The main assumption of our model is that we can use the atmospheric abundances as a trace of the bulk abundances of the giant planets. However, this assumption is certainly under debate \citep{2021ExA...tmp...56H, 2022arXiv220504100G}. If compositional gradients inside of the planet exist, then the atmospheric abundances only represent the minimum abundance of a specific chemical element. Future simulations dedicated to link atmospheric abundances to planet formation need to take detailed interior models into account.

Our planet formation model is of course simplified in many aspects that can influence the planetary abundances, e.g. scattering of small dust grains from the outer gap edges into the planetary feeding zone \citep{2021ApJ...912..107B, 2021MNRAS.506.5969B} or the accretion of smaller pebble sizes once the core has reached the pebble isolation mass for the dominant pebble size \citep{2021arXiv211115218A}, as well as further planetesimal bombardment (e.g. \citealt{2022MNRAS.509..894H}). We also do not include a realistic photoevaporation procedure, which could allow the enhancement of the heavy elements in the disc midplane due to the loss of hydrogen and helium from the disc's upper layers, consequently enriching the gaseous component accreted by the planet resulting in super-solar abundances \citep{2006MNRAS.367L..47G}. In our model, we also do not include the chemical evolution of the gas disc, which can transform the main oxygen and carbon carriers, which also depend on the C/O ratio of the disc \citep{2016arXiv160706710E}, emphasizing the need to constrain stellar abundances to constrain planet formation \citep{2022arXiv220108508R}. For a detailed discussion about limitations of planet formation simulations with respect to atmospheric constraints, see also \citealt{2022arXiv220413714M}.

Our model implicitly implies that WASP-77A b and $\tau$ Bo\"otis b formed further away from their host star and were then scattered inwards. While these scattering events are possible (e.g. \citealt{2008ApJ...686..603J, 2008ApJ...686..621F, 2009ApJ...699L..88R, 2017A&A...598A..70S, 2020A&A...643A..66B}), it implies that other perturbers in the disc are present. These perturbers could either block inward drifting pebbles, thus influencing planetary growth (e.g. \citealt{2015Icar..258..418M, 2019A&A...623A..88B}), and can also have dramatic consequences for the composition of the available solids (e.g. \citealt{2016Icar..267..368M}) and gases \citep{2021A&A...649L...5B, 2021A&A...654A..71S}, requiring a combination of N-body simulations with our here presented planet formation framework to further constrain the formation pathway.

\section{Summary and conclusions}
\label{sec:summary}

In this study we simulated the growth and migration of planetary embryos to gas giants in discs with different $\alpha$-viscosities to derive their atmospheric abundances. The atmospheric abundances are crucially influenced by the planet's formation location and migration history (e.g. \citealt{2011ApJ...743L..16O, 2017MNRAS.469.4102M, 2021A&A...654A..71S}). In line with these studies, our model shows that the atmospheric C/O increases and the corresponding C/H and O/H decrease with increasing planetary formation location. Our results and conclusions apply under the assumption that the atmospheric composition is a tracer of the bulk composition (but see \citealt{2021ExA...tmp...56H, 2022arXiv220504100G}).

Our simulations can explain the observed sub-solar water content of giant planets\footnote{Other planets can also influence the water abundance in the disc, see \citet{2021A&A...649L...5B} and \citet{2021A&A...654A..71S}.}, if the planets form exterior to the water ice line or only migrate across the water ice line very late. On the other hand, planets migrating across the water ice line early feature super-solar water contents, in line with some observed exoplanets as well \citep{2018AJ....155...29W}. The water content of exoplanets, however, needs to be constrainted much better in the future via observations, which show a large range in the water abundance of $\tau$ Bo\"otis b \citep{2021AJ....162...73P, 2022arXiv220514975W}. If planets form exterior to the CO$_2$ evaporation front and only cross the CO$_2$ evaporation front very late in their evolution, our model can reproduce the measured sub-solar C/H and O/H in combination with around solar C/O ratios of WASP-77A b \citep{2021Natur.598..580L}.

Furthermore our model shows that super-solar C/H and O/H ratios are possible when pebble evaporation is taken into account, allowing a match to the atmospheric measurements of $\tau$ Bo\"otis b \citep{2021AJ....162...73P}, if the planet formed exterior to the water evaporation front. This is in contrast to models without pebble evaporation (see appendix~\ref{ap:noevap}), changing the picture put forward by \citet{2011ApJ...743L..16O}, where super-solar C/H and O/H ratios in planetary atmospheres are only possible with additional solid accretion\footnote{This would also enhance the refractory content of giant planet atmospheres, leaving observable signatures \citep{2021A&A...654A..72S}.}.

The effect of pebble evaporation thus allows the formation of planetary atmospheres with sub- and super-solar C/H and O/H. Consequently, our model can match the sub-solar abundances of WASP-77A b \citep{2021Natur.598..580L} and the super-solar abundances of $\tau$ Bo\"otis b \citep{2021AJ....162...73P}, without invoking additional solid accretion for only one of the planets, making our model more general applicable.

Our study further emphasizes that nitrogen could be used as a tracer for the planet formation location \citep{2021ApJ...909...40T, 2021A&A...654A..72S}. This is mostly related to its different chemistry and that most of the nitrogen is stored in the super-volatile N$_2$ component ($T_{\rm evap} = 20$K), reducing the complexity compared to carbon and oxygen bearing molecules. Our model predicts that $\tau$ Bo\"otis b should harbor a super-solar nitrogen abundance, while WASP-77A b's atmosphere should be around solar in nitrogen (Fig.~\ref{fig:owen}). 

Our simulations also show that the effect of pebble evaporation on the composition of planetary atmospheres crucially depends on the disc's viscosity (Fig.~\ref{fig:alpha}). Especially at low viscosities more extreme sub- and super-solar C/H and O/H ratios are possible, due to the pile-up of evaporated material close to the evaporation fronts \citep{2021A&A...654A..71S}, which is diluted faster at higher viscosity. This clearly shows that the $\alpha$ viscosity does not only play a crucial role for planet formation in itself, but also for the atmospheric abundances of planets and allows for a wide range of atmospheric abundances. Furthermore our model implies that a great diversity in atmospheric compositions of hot Jupiters should be expected naturally from their different formation pathways.

\begin{acknowledgements}

B.B., acknowledges the support of the European Research Council (ERC Starting Grant 757448-PAMDORA) and of the DFG priority program SPP 1992 “Exploring the Diversity of Extrasolar Planets (BI 1880/3-1). A.D.S. acknowledges funding from the European Union H2020-MSCA-ITN-2019 under Grant no. 860470 (CHAMELEON). We thank the referee Tristan Guillot for his comments that helped to improve our manuscript.

\end{acknowledgements}

\bibliographystyle{aa}
\bibliography{Stellar}

\begin{thebibliography}{72}
\expandafter\ifx\csname natexlab\endcsname\relax\def\natexlab#1{#1}\fi

\bibitem[{{Alessi} {et~al.}(2020){Alessi}, {Pudritz}, \&
  {Cridland}}]{2020MNRAS.493.1013A}
{Alessi}, M., {Pudritz}, R.~E., \& {Cridland}, A.~J. 2020, \mnras, 493, 1013

\bibitem[{{Alibert} {et~al.}(2018){Alibert}, {Venturini}, {Helled}, {Ataiee},
  {Burn}, {Senecal}, {Benz}, {Mayer}, {Mordasini}, {Quanz}, \&
  {Sch{\"o}nb{\"a}chler}}]{2018NatAs...2..873A}
{Alibert}, Y., {Venturini}, J., {Helled}, R., {et~al.} 2018, Nature Astronomy,
  2, 873

\bibitem[{{Andama} {et~al.}(2022){Andama}, {Ndugu}, {Anguma}, \&
  {Jurua}}]{2021arXiv211115218A}
{Andama}, G., {Ndugu}, N., {Anguma}, S.~K., \& {Jurua}, E. 2022, \mnras, 510,
  1298

\bibitem[{Asplund {et~al.}(2009)Asplund, Grevesse, Sauval, \&
  Scott}]{2009ARA&A..47..481A}
Asplund, M., Grevesse, N., Sauval, A.~J., \& Scott, P. 2009, ARAA, 47, pp.481

\bibitem[{{Baeyens} {et~al.}(2021){Baeyens}, {Decin}, {Carone}, {Venot},
  {Ag{\'u}ndez}, \& {Molli{\`e}re}}]{2021MNRAS.505.5603B}
{Baeyens}, R., {Decin}, L., {Carone}, L., {et~al.} 2021, \mnras, 505, 5603

\bibitem[{{Baumann} \& {Bitsch}(2020)}]{2020arXiv200400874B}
{Baumann}, T. \& {Bitsch}, B. 2020, A\&A, 637, id.A11

\bibitem[{Ben\'{\i}tez-Llambay {et~al.}(2015)Ben\'{\i}tez-Llambay, Masset,
  Koenigsberger, \& Szul{\'a}gyi}]{2015Natur.520...63B}
Ben\'{\i}tez-Llambay, P., Masset, F., Koenigsberger, G., \& Szul{\'a}gyi, J.
  2015, Nature, 520, pp. 63

\bibitem[{{Bergez-Casalou} {et~al.}(2020){Bergez-Casalou}, {Bitsch}, {Pierens},
  {Crida}, \& {Raymond}}]{2020A&A...643A.133B}
{Bergez-Casalou}, C., {Bitsch}, B., {Pierens}, A., {Crida}, A., \& {Raymond},
  S.~N. 2020, \aap, 643, A133

\bibitem[{{Bergin} {et~al.}(2015){Bergin}, {Blake}, {Ciesla}, {Hirschmann}, \&
  {Li}}]{2015PNAS..112.8965B}
{Bergin}, E.~A., {Blake}, G.~A., {Ciesla}, F., {Hirschmann}, M.~M., \& {Li}, J.
  2015, Proceedings of the National Academy of Science, 112, 8965

\bibitem[{{Bi} {et~al.}(2021){Bi}, {Lin}, \& {Dong}}]{2021ApJ...912..107B}
{Bi}, J., {Lin}, M.-K., \& {Dong}, R. 2021, \apj, 912, 107

\bibitem[{{Binkert} {et~al.}(2021){Binkert}, {Szul{\'a}gyi}, \&
  {Birnstiel}}]{2021MNRAS.506.5969B}
{Binkert}, F., {Szul{\'a}gyi}, J., \& {Birnstiel}, T. 2021, \mnras, 506, 5969

\bibitem[{Birnstiel {et~al.}(2012)Birnstiel, Klahr, \&
  Ercolano}]{2012A&A...539A.148B}
Birnstiel, T., Klahr, H., \& Ercolano, B. 2012, A\&A, 539, id.A148

\bibitem[{{Bitsch} {et~al.}(2019){Bitsch}, {Izidoro}, {Johansen}, {Raymond},
  {Morbidelli}, {Lambrechts}, \& {Jacobson}}]{2019A&A...623A..88B}
{Bitsch}, B., {Izidoro}, A., {Johansen}, A., {et~al.} 2019, \aap, 623, A88

\bibitem[{{Bitsch} \& {Johansen}(2017)}]{2017ASSL..445..339B}
{Bitsch}, B. \& {Johansen}, A. 2017, in {Astrophysics and Space Science
  Library}, Vol. 445, {Astrophysics and Space Science Library}, ed. M.~{Pessah}
  \& O.~{Gressel}, 339

\bibitem[{{Bitsch} {et~al.}(2018){Bitsch}, {Morbidelli}, {Johansen}, {Lega},
  {Lambrechts}, \& {Crida}}]{2018arXiv180102341B}
{Bitsch}, B., {Morbidelli}, A., {Johansen}, A., {et~al.} 2018, A\&A, 612,
  id.A30

\bibitem[{{Bitsch} {et~al.}(2021){Bitsch}, {Raymond}, {Buchhave},
  {Bello-Arufe}, {Rathcke}, \& {Schneider}}]{2021A&A...649L...5B}
{Bitsch}, B., {Raymond}, S.~N., {Buchhave}, L.~A., {et~al.} 2021, \aap, 649, L5

\bibitem[{{Bitsch} {et~al.}(2020){Bitsch}, {Trifonov}, \&
  {Izidoro}}]{2020A&A...643A..66B}
{Bitsch}, B., {Trifonov}, T., \& {Izidoro}, A. 2020, \aap, 643, A66

\bibitem[{Booth {et~al.}(2017)Booth, Clarke, Madhusudhan, \&
  Ilee}]{2017MNRAS.469.3994B}
Booth, R.~A., Clarke, C.~J., Madhusudhan, N., \& Ilee, J.~D. 2017, MNRAS, 469,
  p.3994

\bibitem[{{Booth} \& {Ilee}(2019)}]{2019MNRAS.487.3998B}
{Booth}, R.~A. \& {Ilee}, J.~D. 2019, \mnras, 487, 3998

\bibitem[{{Bosman} {et~al.}(2019){Bosman}, {Cridland}, \&
  {Miguel}}]{2019A&A...632L..11B}
{Bosman}, A.~D., {Cridland}, A.~J., \& {Miguel}, Y. 2019, \aap, 632, L11

\bibitem[{Brewer {et~al.}(2017)Brewer, Fischer, \&
  Madhusudhan}]{2017AJ....153...83B}
Brewer, J.~M., Fischer, D.~A., \& Madhusudhan, N. 2017, AJ, 153, id. 83

\bibitem[{{Brogi} \& {Line}(2019)}]{2019AJ....157..114B}
{Brogi}, M. \& {Line}, M.~R. 2019, \aj, 157, 114

\bibitem[{{Col{\'o}n} {et~al.}(2020){Col{\'o}n}, {Kreidberg}, {Welbanks},
  {Line}, {Madhusudhan}, {Beatty}, {Tamburo}, {Stevenson}, {Mandell},
  {Rodriguez}, {Barclay}, {Lopez}, {Stassun}, {Angerhausen}, {Fortney},
  {James}, {Pepper}, {Ahlers}, {Plavchan}, {Awiphan}, {Kotnik}, {McLeod},
  {Murawski}, {Chotani}, {LeBrun}, {Matzko}, {Rea}, {Vidaurri}, {Webster},
  {Williams}, {Cox}, {Tan}, \& {Gilbert}}]{2020AJ....160..280C}
{Col{\'o}n}, K.~D., {Kreidberg}, L., {Welbanks}, L., {et~al.} 2020, \aj, 160,
  280

\bibitem[{{Crida} \& {Morbidelli}(2007)}]{2007MNRAS.377.1324C}
{Crida}, A. \& {Morbidelli}, A. 2007, \mnras, 377, 1324

\bibitem[{{Cridland} {et~al.}(2020){Cridland}, {van Dishoeck}, {Alessi}, \&
  {Pudritz}}]{2020A&A...642A.229C}
{Cridland}, A.~J., {van Dishoeck}, E.~F., {Alessi}, M., \& {Pudritz}, R.~E.
  2020, \aap, 642, A229

\bibitem[{{Debras} \& {Chabrier}(2019)}]{2019ApJ...872..100D}
{Debras}, F. \& {Chabrier}, G. 2019, \apj, 872, 100

\bibitem[{{Eistrup} {et~al.}(2016){Eistrup}, {Walsh}, \& {van
  Dishoeck}}]{2016arXiv160706710E}
{Eistrup}, C., {Walsh}, C., \& {van Dishoeck}, E.~F. 2016, \aap, 595, A83

\bibitem[{{Ford} \& {Rasio}(2008)}]{2008ApJ...686..621F}
{Ford}, E.~B. \& {Rasio}, F.~A. 2008, \apj, 686, 621

\bibitem[{{Giacobbe} {et~al.}(2021){Giacobbe}, {Brogi}, {Gandhi}, {Cubillos},
  {Bonomo}, {Sozzetti}, {Fossati}, {Guilluy}, {Carleo}, {Rainer},
  {Harutyunyan}, {Borsa}, {Pino}, {Nascimbeni}, {Benatti}, {Biazzo},
  {Bignamini}, {Chubb}, {Claudi}, {Cosentino}, {Covino}, {Damasso}, {Desidera},
  {Fiorenzano}, {Ghedina}, {Lanza}, {Leto}, {Maggio}, {Malavolta}, {Maldonado},
  {Micela}, {Molinari}, {Pagano}, {Pedani}, {Piotto}, {Poretti}, {Scandariato},
  {Yurchenko}, {Fantinel}, {Galli}, {Lodi}, {Sanna}, \&
  {Tozzi}}]{2021Natur.592..205G}
{Giacobbe}, P., {Brogi}, M., {Gandhi}, S., {et~al.} 2021, \nat, 592, 205

\bibitem[{{Guilera} {et~al.}(2014){Guilera}, {de El{\'\i}a}, {Brunini}, \&
  {Santamar{\'\i}a}}]{2014A&A...565A..96G}
{Guilera}, O.~M., {de El{\'\i}a}, G.~C., {Brunini}, A., \& {Santamar{\'\i}a},
  P.~J. 2014, \aap, 565, A96

\bibitem[{{Guillot}(2010)}]{2010A&A...520A..27G}
{Guillot}, T. 2010, \aap, 520, A27

\bibitem[{{Guillot} {et~al.}(2022){Guillot}, {Fletcher}, {Helled}, {Ikoma},
  {Line}, \& {Parmentier}}]{2022arXiv220504100G}
{Guillot}, T., {Fletcher}, L.~N., {Helled}, R., {et~al.} 2022, arXiv e-prints,
  arXiv:2205.04100

\bibitem[{Guillot \& Hueso(2006)}]{2006MNRAS.367L..47G}
Guillot, T. \& Hueso, R. 2006, MNRAS, 367, L47

\bibitem[{{Hands} \& {Helled}(2022)}]{2022MNRAS.509..894H}
{Hands}, T.~O. \& {Helled}, R. 2022, \mnras, 509, 894

\bibitem[{{Helled} {et~al.}(2021){Helled}, {Werner}, {Dorn}, {Guillot},
  {Ikoma}, {Ito}, {Kama}, {Lichtenberg}, {Miguel}, {Shorttle}, {Tackley},
  {Valencia}, \& {Vazan}}]{2021ExA...tmp...56H}
{Helled}, R., {Werner}, S., {Dorn}, C., {et~al.} 2021, Experimental Astronomy

\bibitem[{{Ida} \& {Lin}(2008)}]{2008ApJ...673..487I}
{Ida}, S. \& {Lin}, D.~N.~C. 2008, \apj, 673, 487

\bibitem[{Johansen \& Lambrechts(2017)}]{Johansen2017}
Johansen, A. \& Lambrechts, M. 2017, AREP, 45

\bibitem[{{Juri{\'c}} \& {Tremaine}(2008)}]{2008ApJ...686..603J}
{Juri{\'c}}, M. \& {Tremaine}, S. 2008, \apj, 686, 603

\bibitem[{{Kolecki} \& {Wang}(2021)}]{2021arXiv211202031K}
{Kolecki}, J.~R. \& {Wang}, J. 2021, arXiv e-prints, arXiv:2112.02031

\bibitem[{{Lambrechts} \& {Johansen}(2012)}]{2012A&A...544A..32L}
{Lambrechts}, M. \& {Johansen}, A. 2012, A\&A, 544, id.A32

\bibitem[{Lambrechts {et~al.}(2014)Lambrechts, Johansen, \&
  Morbidelli}]{2014A&A...572A..35L}
Lambrechts, M., Johansen, A., \& Morbidelli, A. 2014, A\&A, 572, id. A35

\bibitem[{{Line} {et~al.}(2021){Line}, {Brogi}, {Bean}, {Gandhi}, {Zalesky},
  {Parmentier}, {Smith}, {Mace}, {Mansfield}, {Kempton}, {Fortney}, {Shkolnik},
  {Patience}, {Rauscher}, {D{\'e}sert}, \& {Wardenier}}]{2021Natur.598..580L}
{Line}, M.~R., {Brogi}, M., {Bean}, J.~L., {et~al.} 2021, \nat, 598, 580

\bibitem[{{Lynden-Bell} \& {Pringle}(1974)}]{1974MNRAS.168..603L}
{Lynden-Bell}, D. \& {Pringle}, J.~E. 1974, \mnras, 168, 603

\bibitem[{{MacDonald} \& {Madhusudhan}(2017)}]{2017MNRAS.469.1979M}
{MacDonald}, R.~J. \& {Madhusudhan}, N. 2017, \mnras, 469, 1979

\bibitem[{Madhusudhan {et~al.}(2017)Madhusudhan, Bitsch, Johansen, \&
  Eriksson}]{2017MNRAS.469.4102M}
Madhusudhan, N., Bitsch, B., Johansen, A., \& Eriksson, L. 2017, MNRAS, 469,
  p.4102

\bibitem[{{Mankovich} \& {Fuller}(2021)}]{2021NatAs...5.1103M}
{Mankovich}, C.~R. \& {Fuller}, J. 2021, Nature Astronomy, 5, 1103

\bibitem[{{Miguel} {et~al.}(2022){Miguel}, {Bazot}, {Guillot}, {Howard},
  {Galanti}, {Kaspi}, {Hubbard}, {Militzer}, {Helled}, {Atreya}, {Connerney},
  {Durante}, {Kulowski}, {Lunine}, {Stevenson}, \&
  {Bolton}}]{2022arXiv220301866M}
{Miguel}, Y., {Bazot}, M., {Guillot}, T., {et~al.} 2022, \aap, 662, A18

\bibitem[{{Molli{\`e}re} {et~al.}(2022){Molli{\`e}re}, {Molyarova}, {Bitsch},
  {Henning}, {Schneider}, {Kreidberg}, {Eistrup}, {Burn}, {Nasedkin},
  {Semenov}, {Mordasini}, {Schlecker}, {Schwarz}, {Lacour}, {Nowak}, \&
  {Schulik}}]{2022arXiv220413714M}
{Molli{\`e}re}, P., {Molyarova}, T., {Bitsch}, B., {et~al.} 2022, arXiv
  e-prints, arXiv:2204.13714

\bibitem[{{Molli{\`e}re} {et~al.}(2020){Molli{\`e}re}, {Stolker}, {Lacour},
  {Otten}, {Shangguan}, {Charnay}, {Molyarova}, {Nowak}, {Henning}, {Marleau},
  {Semenov}, {van Dishoeck}, {Eisenhauer}, {Garcia}, {Garcia Lopez}, {Girard},
  {Greenbaum}, {Hinkley}, {Kervella}, {Kreidberg}, {Maire}, {Nasedkin},
  {Pueyo}, {Snellen}, {Vigan}, {Wang}, {de Zeeuw}, \&
  {Zurlo}}]{2020A&A...640A.131M}
{Molli{\`e}re}, P., {Stolker}, T., {Lacour}, S., {et~al.} 2020, \aap, 640, A131

\bibitem[{{Molli{\`e}re} {et~al.}(2015){Molli{\`e}re}, {van Boekel},
  {Dullemond}, {Henning}, \& {Mordasini}}]{2015ApJ...813...47M}
{Molli{\`e}re}, P., {van Boekel}, R., {Dullemond}, C., {Henning}, T., \&
  {Mordasini}, C. 2015, \apj, 813, 47

\bibitem[{Morbidelli {et~al.}(2016)Morbidelli, Bitsch, Crida, Gounelle,
  Guillot, Jacobson, Johansen, Lambrechts, \& Lega}]{2016Icar..267..368M}
Morbidelli, A., Bitsch, B., Crida, A., {et~al.} 2016, Icarus, 267, p. 368

\bibitem[{Morbidelli {et~al.}(2015)Morbidelli, Lambrechts, Jacobson, \&
  Bitsch}]{2015Icar..258..418M}
Morbidelli, A., Lambrechts, M., Jacobson, S.~A., \& Bitsch, B. 2015, Icarus,
  258, p. 418

\bibitem[{{Mordasini} {et~al.}(2009){Mordasini}, {Alibert}, \&
  {Benz}}]{2009A&A...501.1139M}
{Mordasini}, C., {Alibert}, Y., \& {Benz}, W. 2009, A\&A, 1139

\bibitem[{{Mulders} {et~al.}(2018){Mulders}, {Pascucci}, {Apai}, \&
  {Ciesla}}]{2018AJ....156...24M}
{Mulders}, G.~D., {Pascucci}, I., {Apai}, D., \& {Ciesla}, F.~J. 2018, \aj,
  156, 24

\bibitem[{{Ndugu} {et~al.}(2018){Ndugu}, {Bitsch}, \&
  {Jurua}}]{2018MNRAS.474..886N}
{Ndugu}, N., {Bitsch}, B., \& {Jurua}, E. 2018, \mnras, 474, 886

\bibitem[{{Ndugu} {et~al.}(2021){Ndugu}, {Bitsch}, {Morbidelli}, {Crida}, \&
  {Jurua}}]{2021MNRAS.501.2017N}
{Ndugu}, N., {Bitsch}, B., {Morbidelli}, A., {Crida}, A., \& {Jurua}, E. 2021,
  \mnras, 501, 2017

\bibitem[{{\"O}berg {et~al.}(2011){\"O}berg, Murray-Clay, \&
  Bergin}]{2011ApJ...743L..16O}
{\"O}berg, K.~I., Murray-Clay, R.~A., \& Bergin, E. 2011, ApJ, 743

\bibitem[{{{\"O}berg} \& {Wordsworth}(2019)}]{2019AJ....158..194O}
{{\"O}berg}, K.~I. \& {Wordsworth}, R. 2019, \aj, 158, 194

\bibitem[{{Paardekooper} {et~al.}(2011){Paardekooper}, {Baruteau}, \&
  {Kley}}]{2011MNRAS.410..293P}
{Paardekooper}, S.~J., {Baruteau}, C., \& {Kley}, W. 2011, MNRAS, 410, 293

\bibitem[{{Pelletier} {et~al.}(2021){Pelletier}, {Benneke}, {Darveau-Bernier},
  {Boucher}, {Cook}, {Piaulet}, {Coulombe}, {Artigau}, {Lafreni{\`e}re},
  {Delisle}, {Allart}, {Doyon}, {Donati}, {Fouqu{\'e}}, {Moutou}, {Cadieux},
  {Delfosse}, {H{\'e}brard}, {Martins}, {Martioli}, \&
  {Vandal}}]{2021AJ....162...73P}
{Pelletier}, S., {Benneke}, B., {Darveau-Bernier}, A., {et~al.} 2021, \aj, 162,
  73

\bibitem[{{Raymond} {et~al.}(2009){Raymond}, {Armitage}, \&
  {Gorelick}}]{2009ApJ...699L..88R}
{Raymond}, S.~N., {Armitage}, P.~J., \& {Gorelick}, N. 2009, \apjl, 699, L88

\bibitem[{{Reggiani} {et~al.}(2022){Reggiani}, {Schlaufman}, {Healy},
  {Lothringer}, \& {Sing}}]{2022arXiv220108508R}
{Reggiani}, H., {Schlaufman}, K.~C., {Healy}, B.~F., {Lothringer}, J.~D., \&
  {Sing}, D.~K. 2022, \aj, 163, 159

\bibitem[{{Schneider} \& {Bitsch}(2021{\natexlab{a}})}]{2021A&A...654A..71S}
{Schneider}, A.~D. \& {Bitsch}, B. 2021{\natexlab{a}}, \aap, 654, A71

\bibitem[{{Schneider} \& {Bitsch}(2021{\natexlab{b}})}]{2021A&A...654A..72S}
{Schneider}, A.~D. \& {Bitsch}, B. 2021{\natexlab{b}}, \aap, 654, A72

\bibitem[{{Showman} \& {Guillot}(2002)}]{2002A&A...385..166S}
{Showman}, A.~P. \& {Guillot}, T. 2002, \aap, 385, 166

\bibitem[{{Sotiriadis} {et~al.}(2017){Sotiriadis}, {Libert}, {Bitsch}, \&
  {Crida}}]{2017A&A...598A..70S}
{Sotiriadis}, S., {Libert}, A.-S., {Bitsch}, B., \& {Crida}, A. 2017, \aap,
  598, A70

\bibitem[{{Turrini} {et~al.}(2021){Turrini}, {Schisano}, {Fonte}, {Molinari},
  {Politi}, {Fedele}, {Pani{\'c}}, {Kama}, {Changeat}, \&
  {Tinetti}}]{2021ApJ...909...40T}
{Turrini}, D., {Schisano}, E., {Fonte}, S., {et~al.} 2021, \apj, 909, 40

\bibitem[{{Vazan} {et~al.}(2018){Vazan}, {Helled}, \&
  {Guillot}}]{2018A&A...610L..14V}
{Vazan}, A., {Helled}, R., \& {Guillot}, T. 2018, \aap, 610, L14

\bibitem[{{Wahl} {et~al.}(2017){Wahl}, {Hubbard}, {Militzer}, {Guillot},
  {Miguel}, {Movshovitz}, {Kaspi}, {Helled}, {Reese}, {Galanti}, {Levin},
  {Connerney}, \& {Bolton}}]{2017GeoRL..44.4649W}
{Wahl}, S.~M., {Hubbard}, W.~B., {Militzer}, B., {et~al.} 2017, \grl, 44, 4649

\bibitem[{{Wakeford} {et~al.}(2018){Wakeford}, {Sing}, {Deming}, {Lewis},
  {Goyal}, {Wilson}, {Barstow}, {Kataria}, {Drummond}, {Evans}, {Carter},
  {Nikolov}, {Knutson}, {Ballester}, \& {Mandell}}]{2018AJ....155...29W}
{Wakeford}, H.~R., {Sing}, D.~K., {Deming}, D., {et~al.} 2018, \aj, 155, 29

\bibitem[{{Webb} {et~al.}(2022){Webb}, {Gandhi}, {Brogi}, {Birkby}, {de Mooij},
  {Snellen}, \& {Zhang}}]{2022arXiv220514975W}
{Webb}, R.~K., {Gandhi}, S., {Brogi}, M., {et~al.} 2022, \mnras, 514, 4160

\bibitem[{{Welbanks} {et~al.}(2019){Welbanks}, {Madhusudhan}, {Allard},
  {Hubeny}, {Spiegelman}, \& {Leininger}}]{2019ApJ...887L..20W}
{Welbanks}, L., {Madhusudhan}, N., {Allard}, N.~F., {et~al.} 2019, \apjl, 887,
  L20

\end{thebibliography}

\appendix

\section{Model parameters}
\label{ap:parameters}

We show in Table~\ref{tab:parameters} the model parameters used in our study. While we vary the initial planet position and disc viscosity in Section~\ref{sec:alpha}, we keep the other parameters the same. Our model assumes a solar composition of the different initial elemental ratios. We integrate the disc's lifetime until the growing planets have reached the planetary masses of WASP-77A b and $\tau$ Bo\"otis b. This approach implies that the disc conveniently disappears when the final planetary masses are reached and implies a varying disc lifetime depending on the planetary mass that needs to be reached and on the disc's viscosity, because giants grow faster in high viscosity environments. The disc's lifetimes thus vary between 1.5 Myr to 4 Myr. For Fig.~\ref{fig:alpha} we use a fixed lifetime of 3 Myr.

\begin{table}
	\caption{Parameters used throughout this paper.}
	\begin{subtable}{.2\textwidth}
	\vspace{0pt}
	\centering
	\begin{tabular}{c c}
		\hline\hline
		Quantity & Value\\
		\hline
		$a_{p,0}$			 & (3.5,10,30) au\\

		$t_0$			     & 0.05 Myr\\
		$\kappa_\mathrm{env}$& $0.05{{\rm cm}^2/{\rm g}}$\\
		\hline
	\end{tabular}
	\caption{Planet}
	\begin{tabular}{c c}
		\hline\hline
		Quantity & Value\\
		\hline
		$r_\mathrm{in}$       & 0.1 au\\
		$r_\mathrm{out}$      & 1000 au \\
		$N_\mathrm{Grid}$     & 500\\
		\hline
	\end{tabular}
	\caption{Grid}
	\end{subtable}\hfill
	\begin{subtable}{.25\textwidth}
	\vspace{0pt}
		\centering
		\begin{tabular}{c c}
		\hline\hline
		Quantity & Value\\
		\hline
		$\alpha$           & $(1,5,10)\times10^{-4}$\\
		$\alpha_z$         & $10^{-4}$\\
		$M_0$	           & 0.128 $M_\odot$\\
		$R_0$	           & 137 au\\
		$[\mathrm{Fe}/\mathrm{H}]$ & 0.0\\
		$t_\mathrm{evap}$  & 3 Myr\\
		$\epsilon_0$       & $1.24\%$\\
		$u_\mathrm{frag}$  & 5 m/s\\
		\hline
	\end{tabular}
	\caption{Disc}
	\end{subtable}
	\begin{tablenotes}
		\item \textbf{Notes:} Parameters used for the initialization of \texttt{chemcomp} that are used throughout this paper, divided into planetary, numerical and disc parameters. The detailed explanation of these parameters can be found in \citet{2021A&A...654A..71S}.
	\end{tablenotes}

	\label{tab:parameters}
\end{table}

\section{Models without pebble evaporation}
\label{ap:noevap}

Our main model includes the evaporation of inward drifting pebbles that cross ice lines. Here we investigate how the atmospheric composition of planets changes, if the evaporation of inward drifting pebbles is not taken into account. We note that this does not influence the growth and migration of the growing planets \citep{2021A&A...654A..71S}.

\begin{figure}
 \centering
 \includegraphics[scale=1.0]{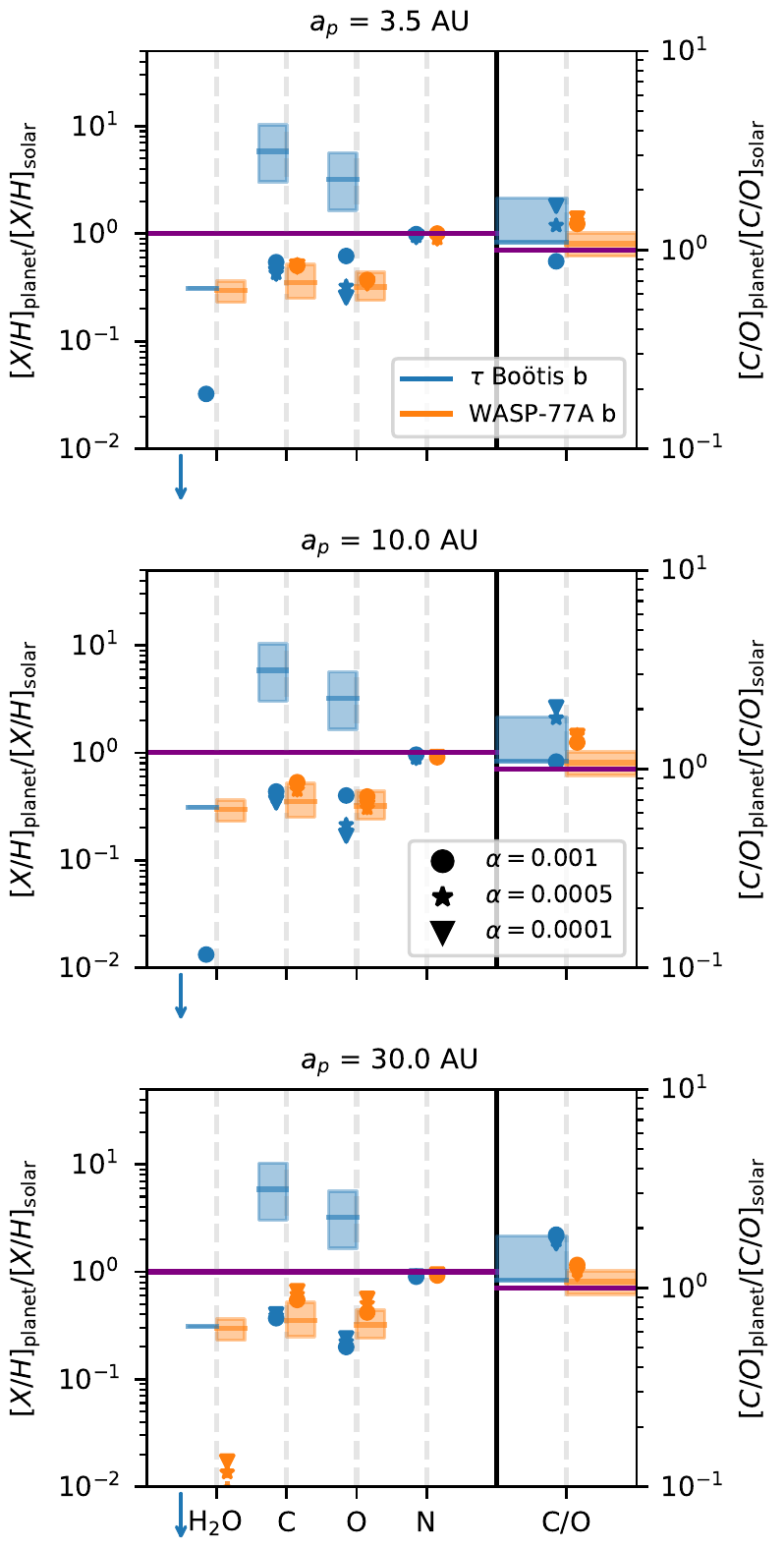}
 \caption{Atmospheric C/H, O/H, N/H, C/O and H$_2$O/H content of planets forming at 3.5, 10 and 30 au in discs with different viscosities. Symbols and color codings are as in Fig.~\ref{fig:owen}. We do not take evaporation of inward drifting pebbles into account, in contrast to Fig.~\ref{fig:owen}, resulting in generally sub-solar abundances.
   \label{fig:owen_ne}
   }
\end{figure}

In Fig.~\ref{fig:owen_ne} we present the atmospheric composition of planets starting at 3.5, 10 and 30 au in discs with different viscosities without taking the evaporation of pebbles into account (compared to Fig.~\ref{fig:owen}, where evaporation is taken into account). As expected from previous models \citep{2011ApJ...743L..16O}, super-solar C/H, O/H and N/H ratios can not be achieved by gas accretion alone, failing to reproduce the atmospheric constraints of $\tau$ Bo\"otis b \citep{2021AJ....162...73P}. Without the contribution of pebble evaporation, super-solar C/H and O/H values can only be achieved if the planet further accretes solids (e.g. planetesimals). This implies that $\tau$ Bo\"otis b would need to accrete further solids to explain its super-solar abundance in contrast to WASP-77A b, opening the question why planetesimal accretion should only be efficient in certain cases. We observe the same effect for the nitrogen abundance, where all planets feature around solar N/H, because the planets migrate across the NH$_3$ evaporation front (where then all nitrogen is in a gaseous component), except if they form in outer regions in low viscosity discs, where 90\% of the nitrogen is in gaseous form. All planets formed in this scenario feature extremely sub-solar water abundances, which could be in line with the measurements of \citet{2021AJ....162...73P} for $\tau$ Bo\"otis b, but are in disagreement with the measurements of \citet{2022arXiv220514975W} for the same planet. However, our formation constraints are derived only from C/H, O/H and C/O, because the water abundances is also heavily influenced by interior processes, not studied in detail in this work.

In contrast, the C/O ratio of the planets can become super-solar, if the planets form exterior to the water ice line. In fact, the model without evaporation allows a match to the atmospheric constraints of WASP-77A b, which feature sub-solar C/H and O/H ratios \citep{2021Natur.598..580L}, leading already to the conclusion that WASP-77A b formed exterior to the water ice line. However, from this model it is unclear how far away from the water ice line WASP-77A b should have formed, giving limited constraints to planet formation (Fig.~\ref{fig:owen_ne}). The recent study by \citet{2022arXiv220108508R} concluded also that WASP-77A b should form beyond the water ice line, but their model does not include pebble evaporation as well. In contrast, if pebble evaporation is included a formation beyond the CO$_2$ snow line is needed, because otherwise too much carbon might be accreted (see Fig.~\ref{fig:owen} and Fig.~\ref{fig:alpha}).

\end{document}